# Exchange coupling in synthetic anion-engineered chromia heterostructures

*Shan Lin,† Zhiwen Wang,† Qinghua Zhang,† Shengru Chen, Qiao Jin, Hongbao Yao, Shuai Xu, Fanqi Meng, Xinmao Yin, Can Wang, Chen Ge, Haizhong Guo, Chi Sin Tang, Andrew T. S. Wee, Lin Gu, Kui-juan Jin,\* Hongxin Yang,\* and Er-Jia Guo\**

S. Lin, Dr. Q. H. Zhang, S. R. Chen, Q. Jin, H. B. Yao, S. Xu, F. Q. Meng, C. Wang, C. Ge, L. Gu, K. J. Jin, and E. J. Guo

Beijing National Laboratory for Condensed Matter Physics and Institute of Physics, Chinese Academy of Sciences, Beijing 100190, China

E-mail: kjjin@iphy.ac.cn and ejguo@iphy.ac.cn

S. Lin and E. J. Guo

Center of Materials Science and Optoelectronics Engineering, University of Chinese Academy of Sciences, Beijing 100049, China

Z. W. Wang and H. X. Yang

Ningbo Institute of Materials Technology & Engineering, Chinese Academy of Sciences, Ningbo 315201, China

E-mail: hongxin.yang@nimte.ac.cn

S. R. Chen, Q. Jin, H. B. Yao, S. Xu, C. Wang, L. Gu, K. J. Jin, and E. J. Guo

School of Physical Sciences, University of Chinese Academy of Sciences, Beijing 100190, China

X. M. Yin

Physics Department, Shanghai Key Laboratory of High Temperature Superconductors, Shanghai University, Shanghai 200444, China

C. Wang, L. Gu, K. J. Jin, and E. J. Guo

Songshan Lake Materials Laboratory, Dongguan, Guangdong 523808, China

H. Z. Guo

School of Physics and Microelectronics, Zhengzhou University, Zhengzhou 450001, China

C. S. Tang

Institute of Materials Research and Engineering, A*STAR, 2 Fusionopolis Way, Singapore, 138634 Singapore

A. T. S. Wee

Department of Physics, National University of Singapore, 2 Science Drive 3, Singapore 117551, Singapore

**Abstract:** Control of magnetic states by external factors has garnered a mainstream status in spintronic research for designing low power consumption and fast-response information storage and processing devices. Previously, magnetic-cation substitution is the conventional means to induce ferromagnetism in an intrinsic antiferromagnet. Theoretically, the anion-doping is proposed to be another effect means to change magnetic ground states. Here we demonstrate the synthesis of high-quality single-phase chromium oxynitride thin films using *in-situ* nitrogen doping. Unlike antiferromagnetic monoanionic chromium oxide and nitride phases, chromium oxynitride exhibits a robust ferromagnetic and insulating state, as demonstrated by the combination of multiple magnetization probes and theoretical calculations. With increasing the nitrogen content, the crystal structure of chromium oxynitride transits from trigonal ($R\bar{3}c$) to tetragonal (4*mm*) phase and its saturation magnetization reduces significantly. Furthermore, we achieve a large and controllable exchange bias field in the chromia heterostructures by synthetic anion engineering. This work reflects the anion engineering in functional oxides towards the potential applications in giant magnetoresistance and tunnelling junctions of modern magnetic sensors and read heads.

**Introduction**

  Magnetoelectric antiferromagnets (AFM), for example archetypical $Cr_2O_3$ (chromia), have been exploited to realize voltage-controlled spintronic devices.[1-5] It is technological important in its pristine form due to two main reasons. One is the Néel temperature ($T_N$) of $Cr_2O_3$ is above room temperature which is suitable for magnetoelectric devices operating in a CMOS environment.[6] Another is its insulating phase can be used as a building block for magnetic tunnelling junction.[7, 8] The observation of robust isothermal electric field mediated exchange bias (EB) effect between exchange coupled $Cr_2O_3$ and ferromagnets heterostructures have stimulated increasing interests.[9-11] The horizontal shift of its magnetic hysteresis loop after high-field cooling below its antiferromagnetic ordering temperature could be used in non-volatile memory units and logic processors. Earlier work has incorporated the ferromagnetic multilayers, such as [Pt/Co][12-14], to investigate the magnetoelectric switching of EB in $Cr_2O_3$ heterostructures. However, the absence of direct coupling between the Néel vector in $Cr_2O_3$ and magnetic field makes the magnetic switching a scientific challenge. He *et al.* reported the uncompensated spins at the surfaces of AFM $Cr_2O_3$ produce a small net magnetization.[15] The EB is only tens of Oersted in heterostructures based on the AFM with uncompensated surfaces. Therefore, extensive efforts have been made to change magnetic ground state of $Cr_2O_3$ via cation substitution to improve the EB field and to enhance $T_N$.[16-19] Substitution in this manner not only introduces chemical strain which influences the magnetic anisotropy and magnetoelastic coupling but may also change the valence state of chromium by creating oxygen vacancies leading to effective spin state transitions. Recently, oxides with mixed-anions become experimentally achievable and the influence of the anion on the physical properties is an emerging research field.[20-25] The exploration of anion doping in the functional oxides may provide an additional degree of freedom in the local anion configurations and even the controllable long-range ordering of anions. Although the ferromagnetic ground state in oxides induced by nitrogen doping has been proposed theoretically, [26-28] the anion-doping effect on the magnetic and structural properties of binary oxides is seldom reported.

  Here we report the successful synthesis of highly-epitaxial $Cr_2O_{3-1.5\delta}N_\delta$ oxynitride thin films using *in-situ* nitrogen doping. X-ray photoemission spectroscopy (XPS) and electron energy loss spectroscopy (EELS) combined with scanning transmission electron microscopy (STEM) are employed as alternative methods to conventional diffraction-based structural characterizations. Second harmonic generation (SHG) measurements demonstrate that the crystal structure of $Cr_2O_{3-1.5\delta}N_\delta$ films transits from trigonal to tetragonal phase with increasing the nitrogen doping level. Using multiple magnetization probes, we reveal that the $Cr_2O_{3-1.5\delta}N_\delta$ films exhibit a robust ferromagnetic ground state and maintain its insulating behavior. This observation is in excellent agreement with our first-principles calculations. Utilizing the anion-engineering, we build a synthetic chromia heterostructure and observe a sizable EB between the AFM $Cr_2O_3$ and ferromagnetic $Cr_2O_{3-1.5\delta}N_\delta$.

**Results**

**First-principles calculations on electronic state of N-doped $Cr_2O_3$**

We firstly tested the structural stability of $Cr_2O_3$ for partial $O^{2-}$ cites occupied by $N^{3-}$ ions. The $Cr_2O_{3-1.5\delta}N_\delta$ oxynitrides preserve the three-dimensional framework instead of the formation of layered structures. To predict the influence of N-doping on the electronic states of $Cr_2O_{3-1.5\delta}N_\delta$, we performed first-principles calculations within the framework of density-functional theory (DFT) as implemented in the Vienna *ab initio* Simulation Package (VASP).[29, 30] We created the theoretical models for $Cr_2O_{3-1.5\delta}N_\delta$ by randomly replacing $O^{2-}$ ions by $N^{3-}$ ions (**Figures 1**a-1d). Four collinear magnetic configurations have been considered (Supplementary Figure S1) and their corresponding energies have been calculated. It is shown that the most stable magnetic configuration of undoped $Cr_2O_3$ is the antiferromagnetic state, whereas the $Cr_2O_{3-1.5\delta}N_\delta$ exhibits stable ferromagnetic states, independent of N doping levels (Table I). Figures 1e–1h show the calculated band structures of $Cr_2O_{3-1.5\delta}N_\delta$ with representative nitrogen concentration $\delta = 0$, 5.56%, 16.67% and 27.78%, respectively. The band dispersions show that indirect bandgaps appear in all $Cr_2O_{3-1.5\delta}N_\delta$, indicating that they all exhibit the intrinsic insulating states. As increasing nitrogen component $\delta$, the bandgap of $Cr_2O_{3-1.5\delta}N_\delta$ decreases, meanwhile the magnetic moments of Cr ions reduce, whereas the moments of N and O ions increase (Figure 1i). The dependence of electronic states on N-doping level agrees with the experimental facts that the pure CrN ($\delta = 2$) is a gapless intrinsic antiferromagnetic metal. Therefore, it is reasonable to observe a bandgap closure in $Cr_2O_{3-1.5\delta}N_\delta$ as increasing $\delta$. Our calculation results suggest that the magnetic states of $Cr_2O_3$ after N doping will transit from antiferromagnetic to ferromagnetic phase and their electronic states maintain insulating with a moderate N doping.

**High-quality $Cr_2O_{3-1.5\delta}N_\delta$ films with controllable nitrogen content**

To verify the theoretical predictions on the physical properties, we fabricated the $Cr_2O_{3-1.5\delta}N_\delta$ thin films on (0001)-oriented $Al_2O_3$ substrates with atomically flat surface using pulsed laser deposition.[31-33] A radio frequency (RF) plasma source was used to generate nitrogen atoms for doping $N^{3-}$ ions into the $Cr_2O_3$ films during the film growth. The RF power and nitrogen flow rate were carefully adjusted to achieve different nitrogen concentrations in the $Cr_2O_{3-1.5\delta}N_\delta$ films. The growth mode of the thin films was monitored by reflection high-energy electron diffraction (RHEED) (Supplementary Figure S2). The streak-shaped diffraction pattern was observed after the deposition, assuring the excellent crystallinity and flat surface of the films. The highly epitaxial growth of as-grown films was further evidenced by the clear Laue oscillations (**Figure 2**a) and narrow rocking curves (Supplementary Figure S2). Figure 2a shows the XRD θ-2θ scans of $Cr_2O_{3-1.5\delta}N_\delta$ thin films around the substrates' (0001) peaks. The film peaks shift toward lower angle with increasing δ. The calculated out-of-plane lattice parameters (*c*) of $Cr_2O_{3-1.5\delta}N_\delta$ thin films increase linearly with the N concentration (inset of Figure 2a). We find that the lattice parameters of $Cr_2O_{3-1.5\delta}N_\delta$ thin films are between those of bulk CrN and bulk $Cr_2O_3$ (Supplementary Figure S3). The *in-situ* nitrogen doping is sensitive to the deposition temperature while is independent of the RF power and gas flow rate (Supplementary Figure S4), suggesting that the nitrogen insertion into $Cr_2O_3$ is a thermal dynamic process. The $Cr_2O_3$ single layers grown in vacuum are nearly stoichiometric with negligible oxygen vacancies because of the binary character and the most stable valence of Cr ions is +3 (Supplementary Figure S5).

Figure 2b shows a representative high-angle annular dark-field (HAADF) image of a $Cr_2O_{2.25}N_{0.5}$ thin film grown on a $Al_2O_3$ substate taken in STEM mode. We do not observe any obvious structural defects or stacking disorders in the $Cr_2O_{2.25}N_{0.5}$ thin films from the STEM wide-range imaging (Supplementary Figure S6). A high-magnified STEM-HAADF image from a selected area in Figure 2b indicates a chemically sharp and coherent interface between $Cr_2O_{2.25}N_{0.5}$ film and $Al_2O_3$ substates. Spatially resolved EELS mapping was performed at the O *K*-, N *K*-, Al *L*-, and Cr *L*-edges, as shown in Figures 2c-2f, respectively. The EELS results suggest that $N^{3-}$ ions are uniformly and randomly distributed within the $Cr_2O_{2.25}N_{0.5}$ film. Annular dark field (ABF) image was acquired simultaneously in the same regime (Supplementary Figure S6). From ABF imaging, we could not well-distinguish the $N^{3-}$ and $O^{2-}$ ions due to the comparable contrasts. Chemical sharpness at the interface and the uniformity within films are also ascertained using EELS profiling (Figure 2g). There is limited chemical intermixing across the interface and the layer sequence is Al-O-Cr-O(N), as indicated by the schematic of atomic arrangements of $Cr_2O_{2.25}N_{0.5}$ and $Al_2O_3$. The relative N and O intensities of EELS reveal that the N/O ratio is ~ 0.2, corresponding to the N doping level δ ~ 0.5. More accurate nitrogen content within $Cr_2O_{3-1.5\delta}N_\delta$ films are determined by depth profiling of each element using ex-situ secondary-ion mass spectrometry (SIMS) (Supplementary Figure S7). The SIMS results confirm that the $N^{3-}$ ions distribute uniformly within films and the concentration of nitrogen increases gradually with N doping level. Quantitatively, we could obtain the nitrogen concentration using the normalized O and N intensities from the stoichiometric $Cr_2O_3$ and CrN single layers, respectively.

Furthermore, we notice that the microstructure of $Cr_2O_{2.25}N_{0.5}$ thin films is in sharp contrast to the corundum-type structure,[34] for example same as the undoped $Cr_2O_3$ films and the $Al_2O_3$ substrate.[35] To quantify the point group symmetry of the $Cr_2O_3$ and $Cr_2O_{2.25}N_{0.5}$ thin films, we performed optical SHG polarimetry measurements. **Figures 3**a and 3b (3c and 3d) show two SHG components $I^{2\omega}_{p-out}$ and $I^{2\omega}_{s-out}$ were taken from a $Cr_2O_3$ ($Cr_2O_{2.25}N_{0.5}$) film, respectively. We find that the best theoretical fits to the data are given by the point group symmetry of $R\bar{3}c$ for the undoped $Cr_2O_3$ films and $4mm$ for the $Cr_2O_{2.25}N_{0.5}$ films, respectively. In both cases, the films are grown on the $Al_2O_3$ substrates. Therefore, we could conclude that the naturally symmetry-breaking surfaces and interfaces are not responsible for the difference in the SHG signals from these films. The SHG symmetry analysis reveals that the crystal structure of $Cr_2O_3$ films transits from the trigonal to tetragonal phase as gradually increasing the N doping. The dramatic structural transition is further confirmed by performing XRD phi-scans and reciprocal space mapping (RSM) (Supplementary Figures S8 and S9). For undoped $Cr_2O_3$ films, a six-fold symmetry is observed in both films and substrates. However, the crystal structure of $Cr_2O_{2.25}N_{0.5}$ films is incommensurate with that of substrates. The six-fold symmetry appears at the (00$l$) peaks, agreeing with the honeycomb-like structure viewed from (111) orientation of the tetragonal/cubic lattices. We attributed the structural transition upon N-doping to the significant increment of the out-of-plane lattice parameters by ~ 5%. Similar symmetry degrading upon single axis elongation has been observed in other oxides or even metals. [36-38] For instance, the lattice structures of $BiFeO_3$ and $PbCoO_3$ single crystals would change from a rhombohedral/cubic phase to a tetragonal phase under a large compressive strain (~ 4.5%) or a uniaxial high-pressure.

**Electronic states and bonding geometries in $Cr_2O_{3-1.5\delta}N_\delta$ thin films**

The electronic states of $Cr_2O_{3-1.5\delta}N_\delta$ thin films were identified by x-ray photoemission spectroscopy (XPS) at room temperature. Before the XPS measurements, the samples' surfaces were cleaned by Ar plasma in vacuum to reduce the influences from the surface impurities and hydroxyls. The XPS measurements were carried out at normal emission with electron taken-off angle 90º relative to the surface plane, yielding to a probe depth of approximately ten nanometers.[39] Figure 4a shows the Cr 2p spectra for three $Cr_2O_{3-1.5\delta}N_\delta$ thin films. The centroids of the Cr $2p_{3/2}$ peaks of all samples are at ~ 576 eV, in good agreement with that for the epitaxial stoichiometric $LaCrO_3$ films,[40] suggesting that Cr ions keep +3 valence state independent of the nitrogen doping. The changes in the main features of Cr 2p spectra with N doping level are clearly marked by two dashed lines. For undoped $Cr_2O_3$ films, the Cr $2p_{3/2}$ peak splits into two separated peaks (α at 576.8 eV and β at 575.6 eV). The occurrence of the peak β at the lower binding energy is attributed to the multiplet splitting, which is a typical feature of $Cr_2O_3$. Quantitatively, we fit the Cr 2p spectra of undoped $Cr_2O_3$ and $Cr_2O_{2.25}N_{0.5}$ films using multiple components (Supplementary Figure S10). As increasing the N doping level from 0.1 to 0.5, the intensity of β peak reduces, indicating the relative percentage of Cr-N binding state increases. **Figures 4**b and 4c present the N 1s and O 1s spectra for the $Cr_2O_{2.85}N_{0.1}$ and $Cr_2O_{2.25}N_{0.5}$ films, respectively. We notice that the O 1s spectra can be composed by two peaks, the peak at 530.1eV is corresponding to the bonding between Cr and O, while the weaker one at 531.6eV is related to the hydroxide/satellite peak. The intensity of N 1s spectra increases significantly with more N doped into films. In contrast, the intensity at the O 1s peak reduces when the partial $O^{2-}$ ions are occupied by $N^{3-}$ ions. The XPS intensities directly reflect the nitrogen concentration in the films, in agreement with SIMS results.

Room-temperature Raman spectroscopy was performed on the $Cr_2O_{3-1.5\delta}N_\delta$ thin films to directly characterize the chemical structure and bonding environment (Supplementary Figure S11). The Raman spectra of $Cr_2O_3$ and $Cr_2O_{2.85}N_{0.1}$ films show similar features, the band located at 553 $cm^{-1}$ is associated with $Cr_2O_3$ compound. Moreover, the broad band at around 665 $cm^{-1}$ in $Cr_2O_{2.25}N_{0.5}$ demonstrate the formation of chemical bonding between $Cr^{3+}$ and $N^{3-}$. Since the CrN does not scatter strongly to the incident light, it is extremely to obtain a sharp Raman shift, compared to the earlier cases.[41-43] However, the intensity of vibration band of undoped $Cr_2O_3$ decreases while the concentration of N increases, illustrating that the crystal structures of the doped films change dramatically from the original corundum structure. These results are consistent with our previous SHG and STEM conclusions.

**Emerging ferromagnetism observed in $Cr_2O_{3-1.5\delta}N_\delta$ films**

The macroscopic magnetic properties of the $Cr_2O_{3-1.5\delta}N_\delta$ films were measured with a superconducting quantum interference device (SQUID). The magnetic field (*H*) dependence of magnetization (*M*) of $Cr_2O_{2.25}N_{0.5}$ and $Cr_2O_{2.85}N_{0.1}$ films is present in Figure 3e. The measurements were performed at 10 K with applying in-plane fields. The results clearly reveal the fact that both samples exhibit clear ferromagnetism. The film with a lower N-doping level has a larger magnetic response. The overall *M* of $Cr_2O_{2.85}N_{0.1}$ is significantly larger than twice that of $Cr_2O_{2.25}N_{0.5}$. Meanwhile, the coercive field ($H_C$) of $Cr_2O_{2.25}N_{0.5}$ is ~ 550 Oe, which is three times larger than that of the $Cr_2O_{2.85}N_{0.1}$. To ensure the intriguing ferromagnetism observed in $Cr_2O_{2.85}N_{0.1}$ films, we performed the synchrotron-based x-ray magnetic circular dichroism (XMCD) spectra at the Cr *L*-edges. Figure 3f shows the x-ray absorption spectra (XAS) near the $L_3$ and $L_2$ edges of Cr under the magnetic fields of ± 1 T at 78 and 300 K.

XMCD is calculated by the difference between two XAS spectra divided by their sum, as shown in Figure 3g. At 78 K, a large XMCD signal with a positive response at the $L_3$ edge and a negative response at the $L_2$ edge implies the intrinsic magnetic moment of the Cr ions and the spin orders are parallel to the applied field. As increasing temperature to 300 K, the XMCD signals disappear, indicating the onset magnetization reduces to zero. The observation of net ferromagnetic order in the $Cr_2O_{3-1.5\delta}N_\delta$ films is surprising since both CrN and $Cr_2O_3$ films are antiferromagnetic. However, these experimental results are consistent with our earlier theoretical calculations and highlight the important role of N-doping in the magnetic phase transition of $Cr_2O_{3-1.5\delta}N_\delta$.

To utilizing the ferromagnetic $Cr_2O_{3-1.5\delta}N_\delta$ thin films, we fabricated the synthetic $[(Cr_2O_3)_N/(Cr_2O_{2.85}N_{0.1})_N]_5$ superlattices (SL), where N (= 5 and 10) denotes the number of unit cells and 5 is the bilayer repetition, by controlling the RF plasma source. The crystal structure, orientation, phase purity, and crystallinity of these SLs were determined by XRD and reflectivity measurements (Supplementary Figure S12). **Figures 5**a and 5b show the schematic and x-ray scattering length density (xSLD) of a representative $[(Cr_2O_3)_{10}/(Cr_2O_{2.85}N_{0.1})_{10}]_5$ SL, respectively. The xSLD depth profile of a SL is derived from a fit to the x-ray reflectivity [44] and describes the chemical composition across the entire SL. We find that the thickness of $Cr_2O_3$ and $Cr_2O_{2.85}N_{0.1}$ layers are approximately 3.2 and 3.0 nm, respectively. Importantly, the atomic density of $Cr_2O_{2.85}N_{0.1}$ layers is larger than that of the $Cr_2O_3$ layers, except for those of the first bilayer due to the different boundary conditions and accommodation of the misfit strains. Increased atomic density after N-doping is attributed to the structural transition induced atomic rearrangement.

Low-temperature field-dependent magnetizations were measured under different cooling fields. Figure 5c shows the representative out-of-plane magnetic hysteresis loops of a $[(Cr_2O_3)_{10}/(Cr_2O_{2.85}N_{0.1})_{10}]_5$ SL at 10 K. After field-cooling from room temperature in the presence of a + 5 T field, a shift of the center of the hysteresis loop along the field axis was observed towards the negative fields. In contrast, on cooling in a –5 T field, the hysteresis loop was biased in the positive direction. This behavior reveals that the typical EB observed at the antiferromagnetic-ferromagnetic heterostructures, with field-shift being a classic signature of unidirectional magnetic anisotropy.[45] The schematic of spin arrangement across the $Cr_2O_3$ and $Cr_2O_{2.85}N_{0.1}$ interface shown in Figure 5a reveals that the spins in the antiferromagnetic $Cr_2O_3$ is pinned by the ferromagnetic $Cr_2O_{2.85}N_{0.1}$ layers under a positive magnetic field. The observation of EB provides key information about the magnetic properties and interactions within the heterostructures. We define the exchange bias field ($H_{ex}$) as the offset of the hysteresis loop along the field axis. Figure 5d summarizes the dependence of EB on temperature for the $[(Cr_2O_3)_N/(Cr_2O_{2.85}N_{0.1})_N]_5$ SLs. Temperature dependent *M-H* loops can be found in Supplementary Figure S13. First of all, EB is maximum for the SL(N = 10) and decreases on reducing the layer thickness to N = 5. Secondly, EB effect decreases when gradually increasing the measuring temperature, finally vanishing in the temperature range of ~ 150 K. Typically, this temperature is the block temperature for the EB effect, suggesting that above this temperature is the lower limit of ferromagnetic $Cr_2O_{2.85}N_{0.1}$ layers. Lastly, we also observe an enhancement of the coercive field ($H_c$) accompanied with EB in SLs (Figure 5e), in agreement with other known EB systems.[46-48] The saturation magnetization ($M_s$) decays with increasing temperature and disappears at around 200-250 K. The $M_s$ for SL(N = 5) is smaller than that for SL(N = 10), which can be attributed to the finite size effect on the magnetization. These features reaffirm the interfacial origin of the exchange coupling as well as the induced ferromagnetism in N-doped $Cr_2O_3$ layers.

**Conclusion**

In summary, we report the theoretical design and experimental observation of the ferromagnetic oxynitride films, namely the N-doped $Cr_2O_3$. The structural and magnetic phase transitions are confirmed to increase the N doping level. A large EB effect is achieved in our precise-controlled growth of $Cr_2O_3$-based superlattices using synthetic anion engineering. This discovery provides a possible strategy to create these unique physical ground states of functional oxides using anions. Without the assistance of other transition metal cations, doping nonmagnetic anions would serve as an alternative avenue for spintronic applications, for instance creating the magnetic domains or manipulating the exchanged magnetic systems. We believe that this work will stimulate further theoretical and experimental studies on other similar binary antiferromagnetic oxides used in ultrafast magnetic memories and modulators.

**Methods**

*Epitaxial thin-film growth:* The $Cr_2O_3$ thin films were fabricated by pulsed laser deposition with a sintered stoichiometric ceramic target. The $Al_2O_3$ substrates were pre-annealed at 1000 ºC for 2 hours to ensure the step-and-terrace feature on the surface. The undoped $Cr_2O_3$ thin films were grown in vacuum at the temperature of 700 ºC and laser fluence of ~ 1.5 J/cm$^2$. For the N-doped $Cr_2O_3$ thin films, the $N^{3-}$ ions were *in-situ* doped into the $Cr_2O_3$ films during the deposition using RF plasma generated atomic nitrogen. The RF power and nitrogen flow rate were adjusted on purpose to achieve different nitrogen doping levels. The homogeneity and nitrogen content were analyzed using second-ion mass spectrometer (SIMS). The [$Cr_2O_3$/$Cr_2O_3$- $_{1.5\delta}N_\delta$] superlattices were fabricated specifically for the magnetic property measurements. The plasma source was shuttered during the growth of $Cr_2O_3$ layers. After the film deposition, all films and superlattices were cooled slowly to room temperature at a rate of -5 °C/min.

***Structural and magnetic characterizations:*** The crystalline qualities and lattice parameters of as-grown thin films were determined by synchrotron-based x-ray diffraction (sXRD) at beamline 1W1A of the Beijing Synchrotron Radiation Facility (BSRF). X-ray reflectivities (XRR) of thin films and superlattices were checked using a laboratory-based four-circle high-resolution x-ray diffractometer (XRD) to ensure the correct layer thickness and bilayer repetitions. The microstructures of thin films were checked at the room temperature using a JEM ARM 200CF scanning transmission electron microscope (STEM) in the high-angle annular dark field (HAADF) mode. Cross-sectional TEM specimens were prepared using a standard focused ion beam (FIB) lift-off process along the [11$\bar{2}$0] orientation. Electron-energy-loss-spectroscopy (EELS) mapping was performed at the N *K*-, O *K*-, Al *L*- and Cr *L*-edges, respectively. The EELS signals were integrated after background subtracting and analyzed using Gatan DigitalMicrograph. The magnetic properties of the thin films and superlattices were measured using a MPMS3 magnetometer (Quantum Design). The magnetic hysteresis loops at each temperature of interest were obtained at a maximum field of ±5 T. The magnetization was calibrated after carefully subtracting linear background to correct the diamagnetic response from the sapphire substrates. The exchange bias measurements on the superlattices were performed at a fix temperature after field-cooling of ±5 T.

***Second harmonic generation (SHG) measurements:*** Optical SHG measurements were performed on a $Cr_2O_3$ and a $Cr_2O_{2.25}N_{0.5}$ films. The SHG signals were taken in the reflection geometry at room temperature. An 800 nm-wavelength laser from a Ti: Sapphire femtosecond laser (Tsunami 3941-X1BB, Spectra-Physics) was used as a pumping beam at an incident angle of 45° with respect to the surface normal. The polarization direction (φ) of the incident light was rotated through an automatically controlled λ/2 wave plate. The second harmonic fields ($E_{2\omega}$) generated through the nonlinear optical process within the films were decomposed into *p*- ($I^{2\omega}_{p-out}$) and *s*-($I^{2\omega}_{S-out}$) polarized components by a polarizing beam splitter. The optical signals were detected by a photon multiplier tube. The SHG polarimetry data were theoretically fitted with analytical models using standard point group symmetries.

***Spectroscopic measurements:*** The in-house x-ray photoemission spectroscopy (XPS) measurements were performed at the room temperature. Spectra were collected at a fixed electron emission angle to compare the valence changes in different thin films. X-ray absorption spectroscopy (XAS) and x-ray magnetic circular dichroism (XMCD) measurements were performed at beamline SINS of the Singapore Synchrotron Light Source (SSLS). All XAS and XMCD measurements were conducted in total electron yield (TEY) mode with a fixed polarity. The incident direction of polarized light is parallel to the surface normal. XMCD measurements were averaged from multiple measurements at magnetic fields of ±1 T applied parallel to the film plane to eliminate the possible nonmagnetic artifacts. We performed XMCD measurements at 78 and 300 K, respectively, to facilitate direct comparison.

*Raman spectroscopy:* A Ti: sapphire laser was used as the excitation light source for the excitation wavelength ranges of 700–910 nm. The excitation spot size was 100 μm in diameter. All low-frequency spectra were collected in the 'depolarized' configuration. The elements for the incoming and scattered light were arranged perpendicular to each other. We measured the as-grown thin films and compared their signal with reference data taken from a sapphire substrate.

*First-principles calculations:* First-principles calculations were performed within the framework of density functional theory (DFT) implemented in VASP. The projected augmented wave (PAW) method was used to describe ion core-electron interactions and exchange correlation effects were treated by the generalized gradient approximation (GGA) in the form of Perdew–Burke–Ernzerhof (PBE) functional. The cutoff energy for plane-wave expansion was set to 520 eV and the Brillouin zone (BZ) was meshed by the 7×7×2 Monkhorst–Pack grids for N-doped $Cr_2O_3$ thin films. To describe the strong on-site Coulomb interaction ($U$) caused by the localized $3d$ electrons in Cr, the GGA + U approximation was employed with an effective U value of 3.5 eV. All calculation parameters are carefully checked to ensure the total energy of the superlattices converged below $10^{-7}$ eV. Atomic positions were fully relaxed with the force acting on each atom being less than 0.001 eV/Å. The optimized unit cell parameters for the $Cr_2O_3$ ($a = b = 5.05$ Å, $c = 13.82$ Å) were chosen as starting point to create the theoretical model for the nitrogen-doped structures. Moreover, the definition of doping concentration (δ) is the ratio of N ions to total nonmagnetic ions, which means δ = 0, 5.56%, 16.67%, and 27.78%.

**Supporting Information**

Supporting Information is available from the Wiley Online Library or from the author.

**Acknowledgements**

S. Lin, Z. W. Wang, and Dr. Q. H. Zhang contributed equally to this work. The authors wish to thank the XPS measurements performed by Dr. Le Wang and Prof. Scott. A. Chambers at Pacific Northwest National Laboratory. This work was supported by the National Key Basic

Research Program of China (Grant Nos. 2019YFA0308500 and 2020YFA0309100), the National Natural Science Foundation of China (Grant No. 11974390, 51672307, 51991344, 52025025, 52072400), the Beijing Nova Program of Science and Technology (Grant No. Z191100001119112), Beijing Natural Science Foundation (Z190010 and 2202060), the Strategic Priority Research Program of Chinese Academy of Sciences (Grant No. XDB33030200). The authors would like to acknowledge the Singapore Synchrotron Light Source (SSLS) for providing the facility necessary for conducting the research. The SSLS is a National Research Infrastructure under the National Research Foundation, Singapore.

Received: ((will be filled in by the editorial staff))
Revised: ((will be filled in by the editorial staff))
Published online: ((will be filled in by the editorial staff))

**Figures and figure captions**

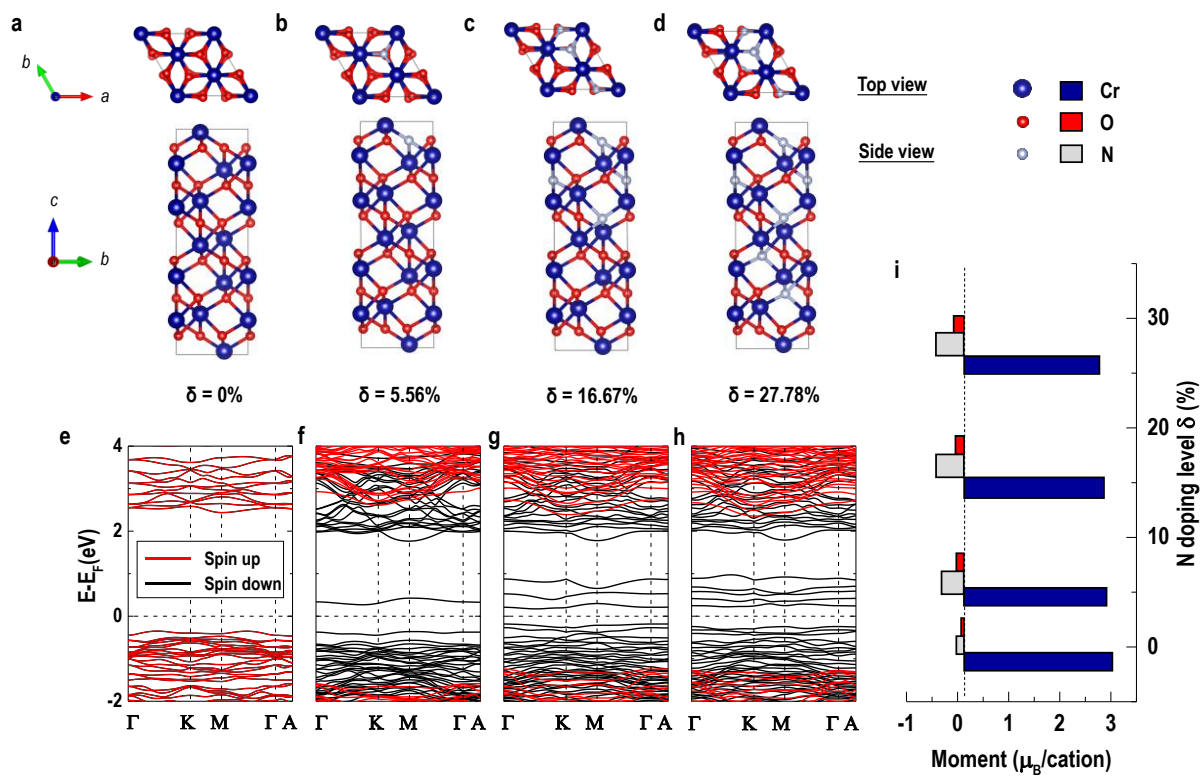

**Figure 1. Evolution of the $Cr_2O_{3-1.5\delta}N_\delta$ band structure with N-doping level**. (a)-(d) Crystal structures and (e)-(h) their corresponding band structures of $Cr_2O_{3-1.5\delta}N_\delta$ with $\delta$ = 0, 5.56%, 16.67%, and 27.78%, respectively. The red and black curves represent the spin-up and spin-down channels of electronic bands, respectively. (i) Calculated elemental specific magnetic moments as a function of the nitrogen content. As the N-doping level increases, the magnetic moment of Cr ions reduces, whereas those of N and O ions increases.

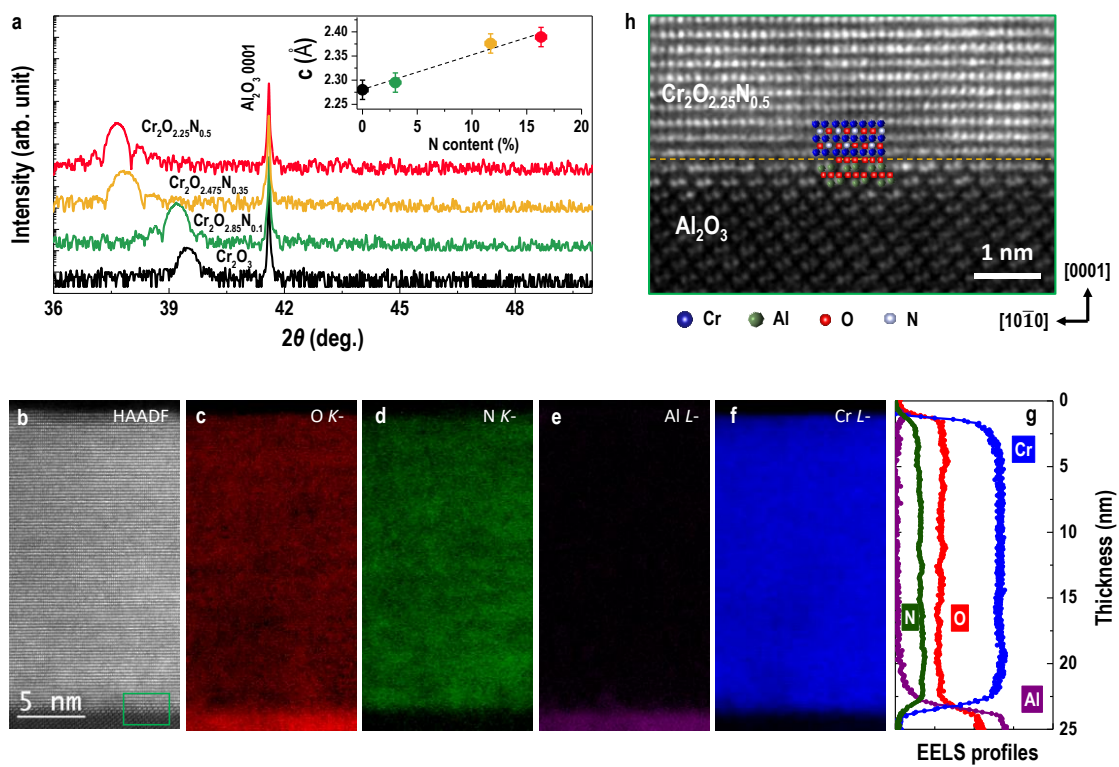

**Figure 2. Structural characterization of $Cr_2O_{3-1.5\delta}N_\delta$ thin films.** (a) XRD θ–2θ scans of a $Cr_2O_3$ film and $Cr_2O_{3-1.5\delta}N_\delta$ films with different N contents. The clear thickness fringes around the films' peaks attributed to the good crystallinities for all samples. Inset: Linear dependence between the out-of-plane lattice constant (*c*) and N content. (b) Scanning transmission electron microscopy (STEM) image of a $Cr_2O_{2.25}N_{0.5}$ film grown on a $Al_2O_3$ substrate acquired using high angle annular dark field (HAADF) mode. The specimen was prepared along the [11$\bar{2}$0] orientation. (h) A high-magnified STEM image from a region marked with the green rectangle in (b) indicates the atomically sharp interface between the film and substrate. Inset of (h): Schematic corresponding crystal structures. The spatial resolved electron-energy-loss-spectroscopy (EELS) maps taken at the (c) O *K*-, (d) N *K*-, (e) Al *L*-, and (f) Cr *L*-edges demonstrate the uniformed distribution of elements and the minimum chemical intermixing across the interface. (g) Elemental line profiles obtained from the EELS mappings averaged across the $Cr_2O_{2.25}N_{0.5}$ films and interfaces.

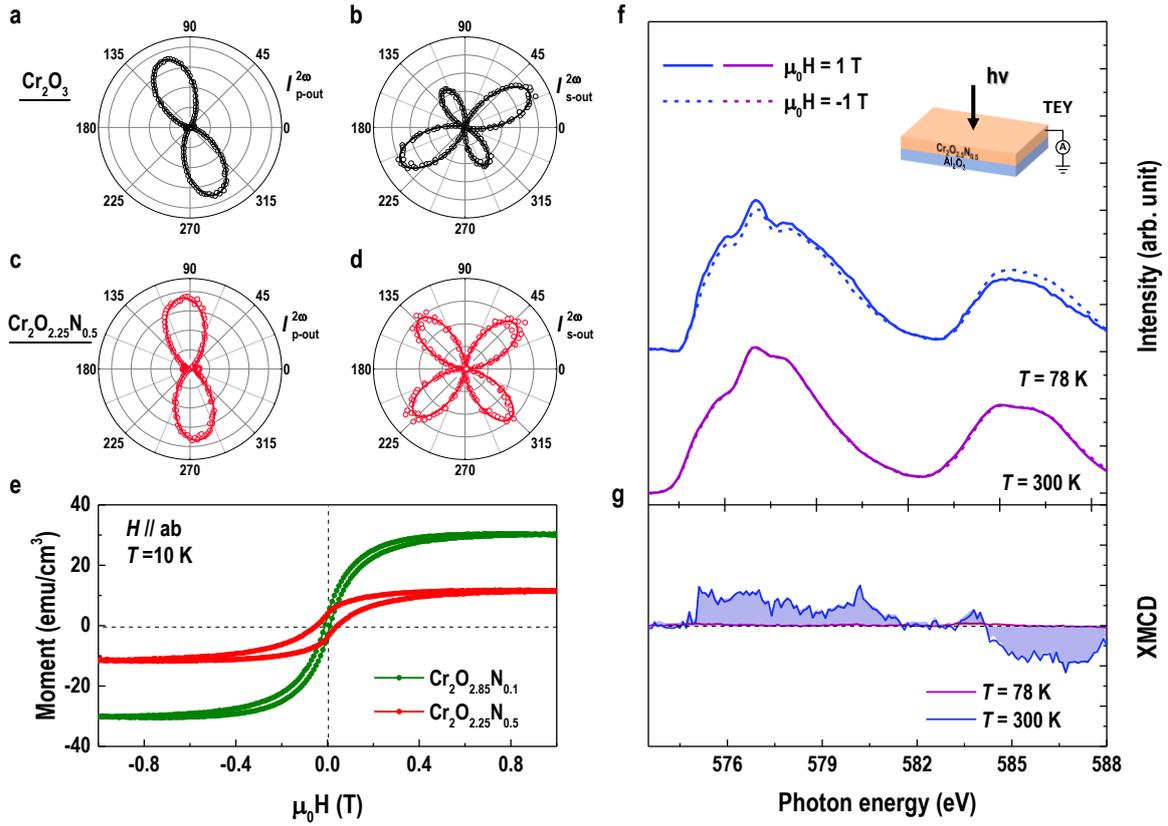

**Figure 3. Observation of ferromagnetism in N-doped $Cr_2O_3$ films.** (a)-(d) Second harmonic generation (SHG) polarimetry from $Cr_2O_3$ and $Cr_2O_{2.25}N_{0.5}$ films. (a) [(c)] $I^{2\omega}_{p-out}$ and (b) [(d)] $I^{2\omega}_{s-out}$ were taken from a $Cr_2O_3$ [$Cr_2O_{2.25}N_{0.5}$] film, respectively. The solid lines in (a)-(d) represent the theoretical fittings to the SHG experimental data. (e) *M-H* curves of $Cr_2O_{2.85}N_{0.1}$ and $Cr_2O_{2.25}N_{0.5}$ films at 10 K. The magnetic fields were applied along the in-plane direction. (f) X-ray absorption spectra (XAS) at Cr *L*-edges were collected from a $Cr_2O_{2.25}N_{0.5}$ film at 78 and 300 K. The measurements were performed in the total electron yield (TEY) mode with applied in-plane fields of ±1 T. The incident photons are parallel to the sample's surface normal. The calculated x-ray magnetic circular dichroism (XMCD) spectra were shown in (g). Non-zero XMCD peak at the Cr *L*-edges is clearly observed at 78 K and the XMCD signal disappears when the measuring temperature rises to 300 K.

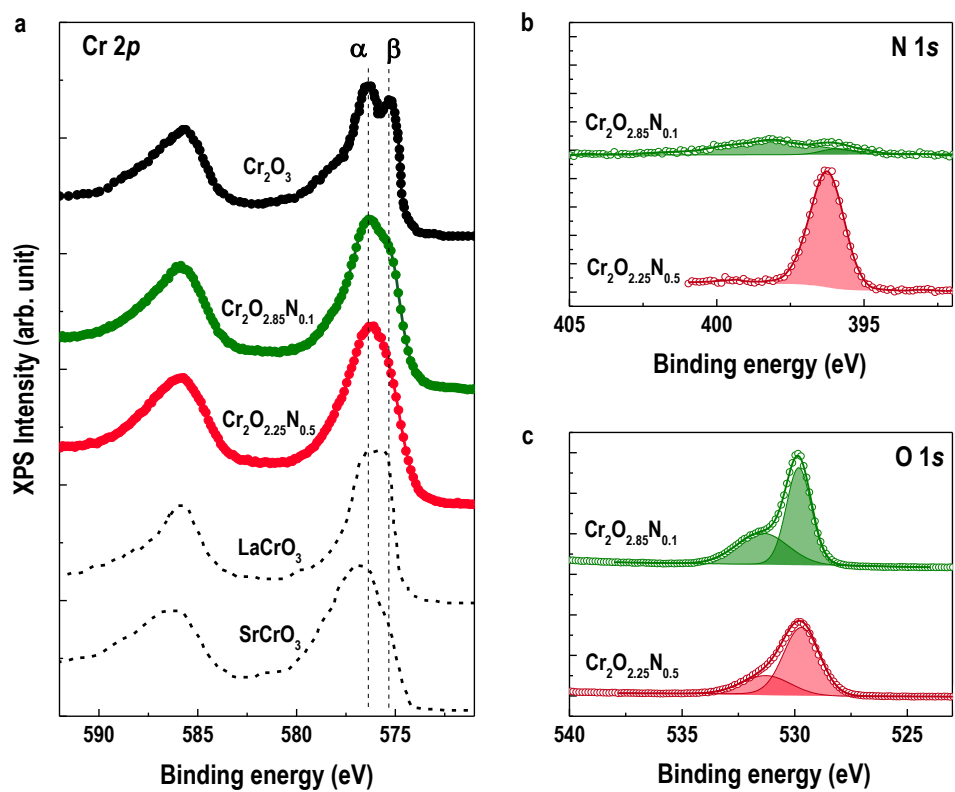

**Figure 4. Electronic states of $Cr_2O_{3-1.5\delta}N_\delta$ thin films.** (a) Cr 2p XPS of $Cr_2O_3$, $Cr_2O_{2.85}N_{0.1}$ and $Cr_2O_{2.25}N_{0.5}$ thin films. The reference spectra (dashed lines) for $Cr^{3+}$ and $Cr^{4+}$ were measured using an oxygen-plasma-annealed $LaCrO_3$ and $SrCrO_3$ films, respectively. The α and β dashed lines denote the binding energy peaks of Cr 2p spectrum from $Cr_2O_3$. As the N content increasing, the intensity of β peak gradually reduces, whereas that of α peak maintains unchanged. (b) O 1s, and (c) N 1s XPS of $Cr_2O_{2.85}N_{0.1}$ and $Cr_2O_{2.25}N_{0.5}$ thin films. The solid lines are the best fits to the experimental data. The colored shadow areas indicate the components of fitting parameters. As increasing the N content, the XPS signal at O 1s decreases, whereas the XPS signal at N 1s increases significantly.

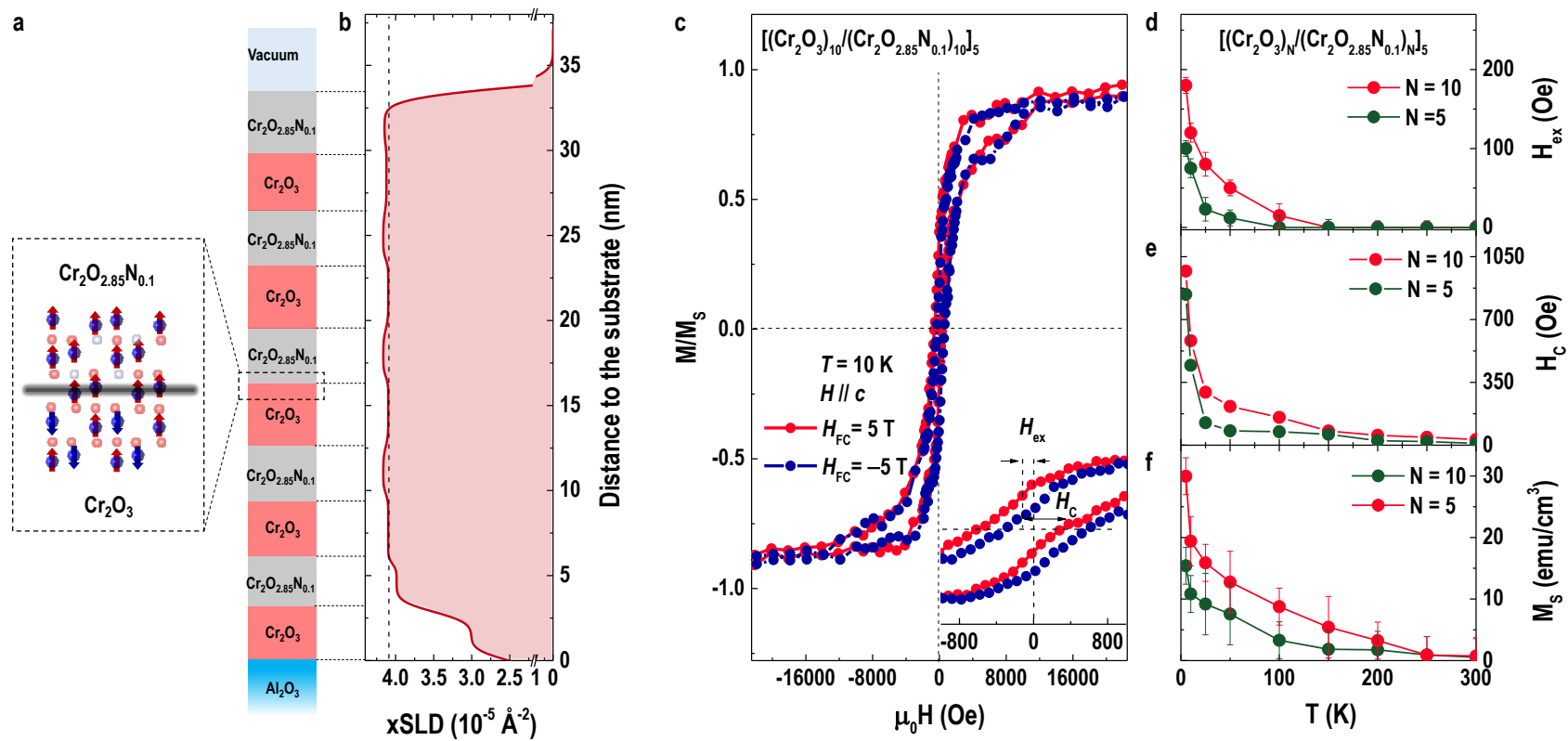

**Figure 5. Magnetic field and temperature dependence of exchange biasing in [(Cr$_2$O$_3$)$_N$/(Cr$_2$O$_{2.85}$N$_{0.1}$)$_N$]$_5$ superlattices**, where N = 5 and 10 representing bilayer repetition. (a) Schematic of sample geometry and spin alignment at the heterointerface under magnetic fields. (b) X-ray scattering length density (xSLD) profile of N=10 superlattices. Dashed line represents the xSLD of bulk Cr$_2$O$_3$. (c) *M-H* hysteresis loops at 10 K for a N=10 superlattice after field-cooling from room temperature in a +5 T field (red circles) and in a -5 T field (blue circles). The magnetic field was applied parallel to the sample's surface normal plane. Inset shows the *M-H* loops at the low field regime. Temperature dependence of (d) exchange bias fields ($H_{ex}$), (e) coercive fields ($H_C$), and (f) saturation magnetization ($M_S$) of superlattices with N = 5 and 10.